\documentclass[12pt,pra,aps,amssymb,amsfonts,amsmath,tightenlines,showpacs]{revtex4}
\usepackage{dsfont}
\usepackage{marvosym}
\usepackage[dvips]{graphicx}
\usepackage{amssymb,amsfonts}

\newcommand{\half}{\mbox{$\textstyle \frac{1}{2}$}}
\newcommand{\re}{\mbox{$\rm e$}}

\newcommand{\rd}{\mbox{$\rm d$}}

\newtheorem{prop}{Proposition}
\newtheorem{definition}{Definition} 

\begin{document}

\title{Thermodynamics of Quantum Heat Bath}

\author{Dorje C. Brody${}$ and Lane P. Hughston${}$}

\affiliation{${}$Department of Mathematics, Brunel University London, Uxbridge UB8 3PH, UK, and \\ 
${}$Department of Optical Physics and Modern Natural Science, 
St Petersburg National Research University of Information Technologies, Mechanics and Optics,\\ 
Kronwerkskii Ave 49,  St Petersburg, 197101, Russia}
\date{\today}

\begin{abstract}
\noindent A model for the thermodynamics of a quantum heat bath is introduced. Under the assumption that the bath molecules have finitely many degrees of freedom and are weakly interacting, we present a general derivation of the equation of state of the bath in the thermodynamic limit. The relation between the  temperature and the specific energy of the bath depends on (i) the spectral properties of the molecules, and (ii) the choice of probability measure on the state space of a representative molecule. The results obtained illustrate how the microscopic features of the molecular constituents determine the macroscopic thermodynamic properties of the bath. Our findings can thus be used to compare the merits of different hypotheses for the equilibrium states of quantum systems. Two examples of plausible choices for the probability measure are considered in detail.  
\end{abstract}


\maketitle
\section{Introduction}
\label{introduction}

\noindent In the early days of quantum mechanics, the notion of a thermal equilibrium state 
for a quantum system was introduced in an \textit{ad hoc} manner, either by  assuming that the 
classical Hamiltonian  in the expression for the partition function should be replaced by 
a set of  discrete energy eigenvalues \cite{AE}, or by the introduction of arguments based on the maximization of entropy  \cite{vN,Jaynes I,Jaynes,Mackey}. More recently, motivated in part by the potential for advances in various quantum 
technologies, a great deal of progress has been made in the modelling of quantum 
equilibrium states \cite{BH1,HT,BH2,BHicqs,BHgqm,JL2000,GM,Bender et al, Naudts, JL,PSW,
GLTZ,BHH2, GLTZ2, Hook,PR,cho,GLMTZ,G,MGE,Fine2, MP,FM} 
and the approach to such equilibrium states  \cite{BH3,LPSW,AS,PDH,RK,RGE, GHT,GE}, in connection with which the analysis of so-called  ``typical'' states \cite{ PSW,GLTZ} for large quantum systems has played a prominent role.
Such studies have led to a better understanding of  the foundations of quantum statistical 
mechanics. But the development of tractable models for quantum equilibrium suitable for the study of phenomena at low temperature has remained elusive. 

In the present paper we address this issue by constructing an exact model for a quantum 
heat bath. The model is well-suited for the analysis of quantum systems at low temperature.  The bath consists of a collection of weakly-interacting components 
(``molecules"), each having finitely-many degrees of freedom. 
We establish that 
in the limit of a large number of such molecules the specific energy of the bath  takes the canonical form 
\begin{eqnarray}
E(\beta)  = \frac { \int_{\mathit\Gamma} H(x) \exp (-\beta H(x) ) \,\mathbb{P} ( {\rm d} x) } 
{  \int_{\mathit\Gamma} \exp (-\beta H(x) )\, \mathbb{P} ( {\rm d} x) },
\label{specific energy} 
\end{eqnarray}
as a function of $\beta = 1/k_BT$, where $T$ is the bath temperature and $k_B$ is Boltzmann's 
constant. Here we write 
\begin{eqnarray}
H(x)= \frac {\langle x|{\hat H}|x\rangle} {\langle x|x\rangle}
\label{energy expectation introduction} 
\end{eqnarray}
for the expectation value of 
the Hamiltonian of a representative bath molecule at a point $x \in {\mathit\Gamma}$ in the space of pure states of the molecule, and $\mathbb{P} ( {\rm d} x)$ denotes a measure on
${\mathit\Gamma}$.  In our approach, the choice of measure on
${\mathit\Gamma}$ is one of the \textit{inputs} of the model. The main result of the paper is to show that the specific entropy of the bath is given by 
\begin{eqnarray}
S(\beta)  = k_B \left [ \beta \,E(\beta) + \log Z(\beta) \right ] 
\label{specific entropy introduction} 
\end{eqnarray}
in the thermodynamic limit, where 
\begin{eqnarray}
Z(\beta)  =  \int_{\mathit\Gamma} \exp (-\beta H(x) )\, \mathbb{P} ( {\rm d} x) .
\label{partition function} 
\end{eqnarray}
In deriving (\ref{specific energy}) and (\ref{specific entropy introduction}), no assumptions are made 
concerning the choice of the measure on the state space of the bath molecules, apart from a   ``completeness" condition to ensure that $\lim_{\beta \to \infty} E(\beta) = E_-$, where $E_-$ denotes the ground state energy of a molecule. 
Thus, we are able to show that the emergence of a canonical equation of state is a generic  feature of an equilibrium aggregation of weakly-interacting finite
quantum systems in the thermodynamic limit. 
Once a choice is made for ${\mathbb P}$, we can work out the 
partition function  of the molecular Hamiltonian under that measure, given by equation (\ref{partition function}),
which in turn allows us to calculate the specific energy and the specific entropy of the bath, and other thermodynamic functions such as the heat capacity and the free energy. 
The results of the analysis can  be used as a basis for the comparison of the merits of different hypotheses for the equilibrium 
states of quantum systems. We examine two candidates for 
${\mathbb P}$ and derive the corresponding expressions for the temperature as a function of 
the specific energy. The first of these is the Dirac measure, which is concentrated at the energy eigenstates 
with even weights, and the other is the uniform measure (Haar measure).
 Both measures satisfy the completeness condition. 

In the case of a bath for which the components are two-level molecules, for instance a paramagnetic 
salt or spin solid such as cerium magnesium nitrate, we show that the Dirac measure leads to 
the familiar relation
\begin{eqnarray}
E(\beta) = \frac {E_1 \, \re^{-\beta E_1} + E_2 \, \re^{-\beta E_2} }
{ \re^{-\beta E_1} +  \re^{-\beta E_2} }
\label{Boltzmann}
\end{eqnarray}
for the specific energy of the bath as a function of the inverse temperature. Here $E_1$ and $E_2$ are the energy levels of the molecular Hamiltonian. 
One can check that $E(\beta)$ is an increasing function of the temperature, and it is evident that if $E_2 > E_1$ then 
$\lim_{\beta \to \infty} E(\beta) = E_1$, and $\lim_{\beta \to 0} E(\beta) = \half (E_1 + E_2)$.  
The form of (\ref{Boltzmann}) is not at first sight surprising, but one should bear in mind that (\ref{Boltzmann}) is a {\it macroscopic}  result, referring to the thermodynamics of the bath as a whole rather than to the behaviour of an individual molecule of the bath. Thus, for example, a sample consisting of one mole of the material of the bath will have energy $N_A \,E(\beta)$ at temperature $T$, where $N_A$ is Avogadro's number. It should be emphasized that we do not at any stage in our analysis assume that the state of the bath can be represented by a Gibbs ensemble, so the result of this example shows that if ${\mathbb P}$ is concentrated on the energy eigenstates then in the limit of a large number of particles we recover the thermodynamics that would be obtained if we had assumed that the bath was in equilibrium with a still larger reservoir held at a constant temperature.    

On the other hand, equation (\ref{Boltzmann}) holds only under the assumption that ${\mathbb P}$ is concentrated at the energy eigenstates. While the usefulness of this assumption as a heuristic tool is indisputable, given the various empirical applications of the resulting formulae, there does not appear to be any generally accepted reason for believing in its universal applicability. 
In fact, it is unreasonable, in the absence of some physical process that would tend to force particles into energy eigenstates (e.g., spontaneous reduction, as described in 
\cite{BHS} and references cited therein), to presume that in a macroscopic sample of a material composed of two-level constituents each of the molecules is necessarily in an energy eigenstate. Such a point of view has been described by Schr\"odinger \cite{schrodinger} as ``indefensible". To be sure, if one were to measure the energy of one of the molecules, then after the measurement the resulting state  would be an energy eigenstate, and in principle if one were to measure the energies of all of the constituents then they would all be in energy eigenstates. 
But  we are not able to make so many measurements at once, and even if we were, there is no reason based on known physics to suppose that such a macroscopic sample will have the property that before the energy measurement each of its constituents will be in an energy eigenstate.  

Nonetheless, an assumption to the effect that each constituent is in an energy eigenstate is implicit in much of the literature of quantum statistical mechanics. In our language this is equivalent to assuming that ${\mathbb P}$ is the Dirac measure. But we do not necessarily make this assumption. An alternative hypothesis is to assume $\textit{a priori}$ that the state of an individual molecule is 
distributed in such a way that the probability of it being in any particular region of the quantum state space 
is proportional to the volume of that region. This corresponds to the situation where 
$\mathbb P$ is the uniform measure on ${\mathit\Gamma}$. In 
the case of a two-level constituent we are then able to work out the resulting formula for the specific energy of the bath as a function of 
the inverse temperature explicitly, leading to the following relation: 
\begin{eqnarray}
E(\beta) = \frac {1} {\beta} + \frac {E_1 \, \re^{-\beta E_1} - E_2 \, \re^{-\beta E_2} }
{ \re^{-\beta E_1} -  \re^{-\beta E_2} }.
\label{uniform}
\end{eqnarray}
One can check that $E(\beta)$ is an increasing function of the temperature, as in the case of the Boltzmann distribution. Likewise one sees  that if $E_2 > E_1$ then 
 $\lim_{\beta \to \infty} E(\beta) = E_1$, as one would expect at zero temperature. The infinite temperature limit is less obvious, but a calculation using l'H\^opital's rule to second order confirms that $\lim_{\beta \to 0} E(\beta) = \half (E_1 + E_2)$, in line with the corresponding result obtained the case of the Dirac measure.  So clearly  (\ref{Boltzmann}) and  (\ref{uniform}) share  features in common. On the other hand, one finds that there are certain important qualitative distinctions between 
 (\ref{Boltzmann}) and  (\ref{uniform}). For example, if we define the heat capacity  as a function of the inverse temperature by setting 
 \begin{eqnarray}
C(\beta) =  - k_B \,  \beta^2 \frac { {\rm d} E(\beta) }  { {\rm d}  \beta} ,
\label{heat capacity}
\end{eqnarray}
then a calculation shows that in the case of the Dirac measure we have 
$\lim_{\beta \to \infty} C(\beta) = 0$ for low temperature, whereas for an aggregation of two-level molecules in the case of the uniform distribution we obtain $\lim_{\beta \to \infty} C(\beta) = k_B$. Indeed, we observe that in the limit of zero temperature  the heat capacity is independent of the value of difference of the two energy levels. The result of the calculation depends only on the fact that the two energy levels are distinct. The existence of a strictly positive ``quantum" of heat capacity as zero temperature is approached can thus be regarded as a hallmark of the uniform distribution.
Whether substances that can be usefully modelled as ideal quantum gases or spin solids having this property can be identified  is of course an open question, but since our theory appears to admit this possibility it would be interesting to pursue the matter.   

\section{Outline of paper}
\label{outline of paper}

\noindent
The remainder of the paper is structured as follows. In Section \ref{Finite-dimensional quantum systems} we summarize those aspects of the geometry of the quantum phase space of a finite dimensional quantum system that we require in the arguments that follow. The space of pure states of such a system has the structure of a symplectic manifold ${\mathit\Gamma}$. 
To model the thermodynamic properties of a composite material for which each of the constituents is taken to be such a system, a measure needs to be introduced on the phase space, which we normalize so it takes the form of a probability measure $\mathbb P$.  The quantum phase space then has the structure of a probability space $({\mathit\Gamma}, \mathcal F, \mathbb P)$.  An important feature of our analysis is that we make it explicit that the state space of a quantum system should have such a structure. The most natural choice of 
$\mathbb P$ is the uniform measure. This choice is natural in the sense that it relies for its specification on no structure other than that present in any finite dimensional quantum system. 

If further structure is introduced, then the choice of natural measures widens. In particular, if ${\mathit\Gamma}$ is given the structure of a Hamiltonian system with the specification of a Hamiltonian function $H : {\mathit\Gamma} \to {\mathds R}$, then the Hamiltonian function can be used to construct alternative measures on ${\mathit\Gamma}$. In principle, any choice of Hamiltonian function on ${\mathit\Gamma}$ can be made, but we adhere to the view that naturality requires that the choice of Hamiltonian function should be dictated by the consideration of structures that are essential to the physics of the situation, without the introduction of extraneous constructions. 
If the Hamiltonian operator of the finite dimensional quantum system is specified, then this operator can be used to construct a Hamiltonian function on ${\mathit\Gamma}$, given by the expectation value of the Hamiltonian operator at each point of the phase space. The resulting Hamiltonian function can be interpreted as a {\it random variable} on the probability space 
$({\mathit\Gamma}, \mathcal F, \mathbb P)$.

We take a set-theoretic approach to the introduction of thermodynamic notions. The  ideas are laid out in Section \ref{Entropy of a subspace of a state space}. The essentially probabilistic approach to quantum statistical mechanics that we outline here offers the basis of a new formulation of the principles of thermodynamics. The entropy of a measurable subset 
$A \subset ({\mathit\Gamma}, \mathcal F, \mathbb P)$ of a quantum phase space is taken 
in Definition \ref{entropy definition} to be given by
$S[A] = k_B \log \mathbb P(A)$. This is consistent with the intuition that the points of ${\mathit\Gamma}$ represent the possible ``microstates"  of the system, and that $\mathbb P(A)$ acts as a measure of the ``number" of such microstates.  In fact, the interpretation of the entropy is clearer in the present context than it is in the usual setup. This is on account of the fact that we give the quantum state space the structure of a probability space. Various entropies can be defined, and the entropy associated with a physical situation is the one for which the constraints on the phase space appropriate to that situation are satisfied. This applies in particular when we extend the definition to composite systems. 

In Sections \ref{construction of quantum heat bath}--\ref{Law of Large Numbers} we proceed in this spirit to model a quantum heat bath by taking the random variable representing the total energy of the bath to be the sum of a large number of independent identically distributed random variables representing the energies of the molecules of the bath. It is assumed that the admissible states of the bath are concentrated on the subspace of the total state space of the bath for which the bath particles are disentangled. This is what one means by an ideal gas of ``weakly interacting" molecules. The entropy of a measurable subset of the space of disentangled states of $n$ molecules is defined  with the understanding that $\mathbb P$ is a product measure on the product space of $n$ copies of the state space of an individual bath molecule.  The specific entropy of the bath at specific energy $E$ is then defined in the thermodynamic limit, providing this limit exists,  by 
 \begin{eqnarray}
S(E) =  \lim_{n \to \infty} \frac {1}{n}\, k_B 
\log \mathbb P\left[ \frac {1}{n} \sum_{j= 1}^{n} H_j \leq E\right].
\label{bath entropy}
\end{eqnarray}
Here the $H_j$ ($j = 1, 2, 3, \dots, n$) are the Hamiltonian functions associated with the various molecules. To show that the limit exists in the sense that the argument on the right hand side of  (\ref{bath entropy}) converges for large $n$ to a definite value, which ishould not be regarded as  {\it a priori} obvious, we require that $\mathbb P$ should satisfy a certain completeness condition that ensures that for each possible value of the specific energy of the bath there exists a temperature at which that energy can be reached. More specifically, let $E_-$ denote the lowest eigenvalue of the Hamiltonian of a bath molecule and write
\begin{eqnarray}
\bar E =   \int_{{\mathit\Gamma}} H(x)  \,\mathbb{P} ( {\rm d} x) 
\label{H bar} 
\end{eqnarray}
for the average value of $H(x)$ under $\mathbb{P}$.
In Definition  \ref{H complete} we introduce the required notion of completeness. We say that $\mathbb{P}$ is ``complete" (for the given Hamiltonian) if  
$\lim_{\beta \to \infty} \,E(\beta) = E_-$ and if for any $\epsilon \in (E_-, \bar E] \subset {\mathds R}$ there exists a unique value of $\beta \in {\mathds R}^+$ such that $\epsilon= E(\beta)$, where the function $E(\beta)$ is defined by equation (\ref{specific energy}). 
We show in Proposition \ref{completeness proposition} that a sufficient condition for   $\mathbb{P}$ to be complete is that for any choice of $\epsilon > E_-$ it holds that 
\begin{eqnarray}
{  \int_{{\mathit\Gamma}} \mathds{1} \{H(x) < \epsilon\} \mathbb{P} \, ( {\rm d} x) } > 0,
\label{completeness condition} 
\end{eqnarray}
where $\mathds{1} \{A\}$ denotes the indicator function for the set $A$, so $\mathds{1} \{H(x) < \epsilon\} = 1$ for $x$ such that $H(x) < \epsilon$ and 
$\mathds{1} \{H(x) < \epsilon\} = 0$ for $x$ such that $H(x) \geq \epsilon$. Intuitively, this condition means that $\mathbb{P}$ has to be sufficiently well spread over $\mathit\Gamma$.

The main result of the paper is summarized in Proposition \ref{fundamental thermodynamic relation}, where we prove that the thermodynamic limit (\ref{bath entropy}) exists under the completeness condition, and is given by 
 \begin{eqnarray}
S(E) =  k_B \left[  \, \beta(E) \, E +  \log Z(\beta(E)) \, \right].
\label{entropy formula}
\end{eqnarray}
Here, for each value of $E \in (E_-, \bar E]$ the corresponding value of the inverse temperature $\beta(E)$ is the unique solution of equation (\ref{specific energy}), and the partition function $Z(\beta(E))$ is defined by (\ref{partition function}).

Note that the specific entropy and the temperature of the bath are completely determined as functions of the specific energy by formulae involving calculations on the state space of a representative molecule. Thus a more or less complete solution of the problem of the identification of the macroscopic equation of state of the bath substance can be given in terms of quantities that are determined at a microscopic level. 

In Section \ref{Independence of width of energy shell} we show in 
Proposition \ref{energy shell prop} that the specific entropy is concentrated in a narrow band of specific energies just at and beneath $E$. That is to say, for any strictly positive value of $\Delta$ the specific entropy of the ``energy shell"  $[E - \Delta, E]$ defined by 
 \begin{eqnarray}
S[E - \Delta, E]=  \lim_{n \to \infty} \frac {1}{n} \, k_B
\log \mathbb P\left[ E - \Delta \leq \frac {1}{n} \sum_{j= 1}^{n} H_j \leq E\right]
\label{bath entropy with band}
\end{eqnarray}
is independent of  $\Delta$ and is equal to the value of $S(E)$ given by (\ref{entropy formula}). 
In Section  \ref{examples} we conclude with 
some examples. In the situation where we assume that the bath constituents are in energy eigenstates, the results of our theory reduce to those of the usual more heuristic approach to quantum statistical mechanics. Our results, however, are obtained under minimal assumptions, starting from Definition \ref{entropy definition}, and are obtained by taking the thermodynamic limit, the existence of which we establish as a fact. For other choices of 
$\mathbb{P}$ the results that follow differ from those of the usual approach, and thus offer the prospect of new modelling methods as well as ways of testing the theory. Our analysis as it stands is applicable to 
weakly-interacting quantum systems, and it remains to be 
seen whether a theory of strongly-interacting quantum systems can be formulated along 
similar lines. 

\section{Finite-dimensional quantum systems}
\label{Finite-dimensional quantum systems}

\noindent
The present work is motivated in part by our growing need to understand the role of thermodynamic effects in the development of quantum technologies. One wishes to establish a consistent theory of the thermodynamics of ``quantum machines", and to understand the limitations on the efficiencies of quantum computations imposed by thermal interference.  In practice this means the development of quantum statistical mechanics in a setting where the constituents of the systems under analysis can be represented by finite dimensional Hilbert spaces. Here we consider the quantum thermodynamics of a quantum heat bath. The heat bath is assumed to consist of a large number of weakly interacting ``molecules", each of which has finitely many degrees of freedom. We assume that the molecular interactions 
are sufficiently weak to ensure that the support of the state space of the bath is the topological 
product of the projective Hilbert spaces of the individual molecules.  It is an essential feature 
of the notion of a collection of weakly interacting molecules that there should be negligible 
entanglement between the states of the various molecules, and that the energies associated 
with the interactions between the various molecules can be neglected. 
Our goal is to set up the problem in such a way that the properties of the quantum heat bath can be calculated explicitly in the limit as the number of constituents of the heat bath is taken to be large.

We begin by making some comments about finite quantum systems. There are three ingredients required for the representation of a finite system
in quantum theory. These are:  (i) the state space of the system, denoted 
${\mathit\Gamma}$,  (ii) the system Hamiltonian $\hat H$, and (iii) a normalized 
measure ${\mathbb P}$ on ${\mathit\Gamma}$, which determines how ``averages" are taken over ${\mathit\Gamma}$. 
We take the system to be represented by a Hilbert 
space ${\mathcal H}$ of  finite dimension $r$. The state space (or ``phase 
space") of the system is the complex projective space ${\mathit\Gamma} = 
\mathds{CP}^{r-1}$ given by the space of rays through the origin in 
${\mathcal H}$.   The pair $({\mathit\Gamma}, \mathcal F)$ is then a measurable space, where $\mathcal F$ denotes the Borel sigma-algebra  generated by the open sets of ${\mathit\Gamma}$. 
The use of the term ``phase space" in the present context is justified by the fact that 
${\mathit\Gamma}$ has a natural symplectic structure (see \cite{BHgqm} and references cited therein). 

From the operator $\hat H$ one is able to construct an associated Hamiltonian function on the state space
$ {\mathit\Gamma}$, given for each point $x \in {\mathit\Gamma}$ by the expectation value of $\hat H$ in the corresponding state, and we write 
$H(x)= {\rm tr} [\hat H \,  \hat\Pi(x)]$, 
where $\hat \Pi(x) = |x\rangle \langle x| / \langle x|x\rangle $
is the projection operator on to the element $|x\rangle \in {\mathcal H}$ corresponding to the point $x \in {\mathit\Gamma}$. 
The significance of the Hamiltonian function is that the associated Hamiltonian vector field on ${\mathit\Gamma}$ obtained by taking the symplectic gradient of $H(x)$ generates the 
Schr\"odinger trajectories of quantum mechanics. As far as we are aware, the suggestion that the resulting phase space structure for quantum theory could be used as a basis for quantum statistical mechanics first appears in 
\cite{BH1, BH2}, though elements of the approach can be found in works of earlier authors, and in this connection we mention Khinchin's treatise \cite{K2} and the posthumously published notes of Bloch \cite{Bloch}.   
The point is that the symplectic manifold
${\mathit\Gamma}$, equipped with the Hamiltonian function $H(x)$, has the phase-space structure required for systematic use of the mathematical methods of ``classical" statistical mechanics alongside modern probability theory, and hence offers a basis for a logical development of the subject. 

The choice of \textit{a priori} measure ${\mathbb P}$ on $({\mathit\Gamma}, \mathcal F)$ is not fixed in advance, except to the extent that it must be natural to the physical problem under consideration, which for equilibrium typically means either (a) the uniform distribution (where the measure of a set is the volume of that set under the unitary invariant Fubini-Study metric on ${\mathit\Gamma}$) or (b) a distribution associated with the Hamiltonian. We shall take the view that this is a modelling choice, and that the merits of any particular choice of measure can be judged by its usefulness in a specific context. There is no requirement in case (b) that the measure should be absolutely continuous with respect to the uniform measure (and hence related to it via a strictly positive density function), but it is tempting to hypothesize that this should be the case, notwithstanding the fact that for some purposes (including common applications) a discontinuous distribution, concentrated at the eigenstates of the Hamiltonian, is useful.  

In the case of a finite dimensional system one further assumption can be made without loss of generality, and this is that the measure can be normalized in such a way that the total measure of the phase space is unity. Thus if we write ${\mathbb P(A)}$ for the measure of any measurable set $A \in \mathcal F$, then ${\mathbb P({\mathit\Gamma})} = 1$. With this convention, the quantum phase space has the structure of a probability space $({\mathit\Gamma}, \mathcal F, {\mathbb P})$, upon which the Hamiltonian function 
$H: {\mathit\Gamma} \rightarrow {\mathds R}$ is a random variable.  This means we can use the tools of probability theory for the solution of problems in quantum statistical mechanics, e.g., for the computation of  averages in the thermodynamic limit. 

The interpretation of phase-space functions as random variables is a feature of the quantum theory of finite systems, and as such offers an advantage over the situation in classical statistical mechanics, where this interpretation is generally not available \cite{K1}. It is important to note, however,  that we have no need in our approach to assign any significance to the term ``random state" beyond the fact that it means a point in a state space equipped with a probability measure. In particular, we do not rely in any special way on the theory of measurement in quantum mechanics. Nor, unlike the recent literature on typical states, do we make use of the notion of ``choosing a state at random". Likewise, no Bayesian reasoning is involved beyond the idea of introducing an {\it a priori} distribution on the space of states.  
On the other hand, the use of modern probabilistic methods as a basis for a critical re-examination of foundational issues in the quantum statistical mechanics of finite dimensional systems is indeed both suggestive and essential. 

\section{Entropy of a subspace of a state space}
\label{Entropy of a subspace of a state space}

\noindent
We take the view that the entropy of a quantum system can be expressed as a function of the number of microstates accessible to it. This suggests: 
\begin{definition} 
\label{entropy definition}
The entropy associated with a measurable subset $A \subset {\mathit\Gamma}$ of a quantum phase space $({\mathit\Gamma}, \mathcal F, {\mathbb P})$ with measure $\mathbb P$ is given by 
\begin{eqnarray}
S[A] = k_B \log \mathbb P(A). 
\end{eqnarray}
\end{definition} 
It should be evident by our conventions that $S[{\mathit\Gamma}] = 0$ and that if $A$ is a proper subset of ${\mathit\Gamma}$ then $S[A] < 0$. If $A$ is a set of measure zero, then $S[A] = - \infty$. It will be convenient therefore to define the entropy as a map 
\begin{eqnarray}
S : \mathcal F \to {\mathds R}^-\cup\{-\infty\}. 
\end{eqnarray}
Then, since $\mathbb P(A)+ \mathbb P(B) \geq \mathbb P(A) \mathbb P(B) $ for all $A, B \in \mathcal F$, for $A$ and $B$ disjoint we have 
\begin{eqnarray}
S[A\cup B]\geq S[A] + S[B ] . 
\end{eqnarray}

In what follows we need to consider the entropies associated with multi-particle systems. The relevant ideas can be conveyed by giving an example. We consider the state space of a two-particle system, when each of the particles is a two-level system. Then the Hilbert spaces of the individual particles are two-dimensional, and the Hilbert space of the composite system is four dimensional. The space of pure states of the composite system is $\mathds{CP}^3$, and the space of disentangled states is a quadric surface 
$\mathcal Q = \mathds{CP}^1 \times \mathds{CP}^1  \subset \mathds{CP}^3$. If we endow 
$\mathds{CP}^3$ with the uniform measure (or any measure absolutely continuous with the uniform measure) then the quadric $\mathcal Q$ will have measure zero, as will any subset of $\mathcal Q$. 
In applications, however, we need to consider systems that are disentangled. For example, in our model for a heat bath we consider a system of molecules with the property that the states of the molecules are mutually disentangled. In the case of a two-particle composite this corresponds to the situation where the measure on 
$\mathds{CP}^3$ is concentrated on $\mathcal Q$. Then $\mathcal Q$ has measure unity, and any measurable subset of $\mathds{CP}^3$ that has a null intersection with $\mathcal Q$ has measure zero. 

Now, the measurable subsets of $\mathcal Q$ can be rather complicated. Suppose, for example, that the measure on $\mathcal Q$ is taken to be the product measure given by the product of the uniform measures on each of the two 
$\mathds{CP}^1$s.  Clearly if $C_1$ and $C_2$ are measurable subsets of the first 
$\mathds{CP}^1$ and the second $\mathds{CP}^1$ respectively, then $C_1 \times C_2$ is a measurable subset of $\mathcal Q$. But any countably additive union or intersection of subsets of this type will also be a measurable subset of $\mathcal Q$. Let us write 
$\mathds{CP}_{(1)}^1$ and $\mathds{CP}_{(2)}^1$ for the two $\mathds{CP}^1$s. 
We shall be interested in measurable subsets of $\mathcal Q$ of the type 
\begin{eqnarray}
A_1 = C_1 \times \mathds{CP}_{(2)}^1 \quad {\rm and} \quad  
A_2 = \mathds{CP}_{(1)}^1 \times C_2, 
\label{disentangled states}
\end{eqnarray}
where $C_1 \subset \mathds{CP}_{(1)}^1$ and $C_2 \subset \mathds{CP}_{(2)}^1$. 
Then the measure of the set $A_1$ has the interpretation of being the probability that the first particle is in $C_1$ and the second particle is anywhere in $\mathds{CP}_{(2)}^1$, whereas the measure of the set $A_2$ has the interpretation of being the probability that the second particle is in $C_2$ and the first particle is anywhere in $\mathds{CP}_{(1)}^1$. When we speak of a particle being in such-and-such location, we are of course referring to the state of the particle being in this location; but it is convenient to make use of the physical language of location in phase space. The point here is that the ``events" $A_1$ and $A_1$ thus defined are \textit {independent} under  the product measure on $\mathcal Q$. One sees that $A_1 \cap A_1 = C_1 \times C_2$, and thus
\begin{eqnarray}
\mathbb P( A_1 \cap A_2) = \mathbb P( A_1 ) \, \mathbb P( A_2). 
\end{eqnarray}
This implies that  the entropies associated with measurable sets of the form (\ref{disentangled states}) have the property that 
\begin{eqnarray}
S[A_1 \cap A_2] = S[A_1] + S[A_2 ]. 
\end{eqnarray}
It should be evident that this additivity structure of our entropy generalizes straightforwardly 
to the case of $n$-particle systems. 

\section{Construction of quantum heat bath}
\label{construction of quantum heat bath}

\noindent
Now suppose we consider a quantum heat bath $B$ consisting of $n$ molecules, all of the same type for simplicity. Each molecule will be represented by a finite quantum system for which the Hilbert space is of some dimension $r$. Let us write 
\begin{eqnarray}
H_B^{(n)}(x) = H_1(x) + H_2(x) +\cdots + H_n(x) 
\label{bath Hamiltonian}
\end{eqnarray}
for the total Hamiltonian function of the bath, where  $H_j(x)$, $j = 1, 2, \dots, n$, are the Hamiltonian functions of the various bath molecules. Here we write
\begin{eqnarray}
H_j(x)=\frac {\langle x|{\hat H}_j|x\rangle} {\langle x|x\rangle} 
\end{eqnarray}
for the expectation of the Hamiltonian ${\hat H}_j$ ($j=1, 2, \dots, n$) of molecule $j$
in the bath state $|x\rangle$. 
It follows from the assumed absence of entanglement among the bath molecules that a factorization of the form
\begin{eqnarray}
|x\rangle = |x_1\rangle |x_2\rangle \cdots |x_n\rangle 
\end{eqnarray}
holds, and hence that for each $j$ the  Hamiltonian function  $H_j(x)$ depends only on the state space variable $x_j$ associated with molecule $j$.

We shall assume that the state space of the bath is endowed with a probability measure $\mathbb P$ concentrated on the Segre variety of disentangled states 
\begin{eqnarray}
\mathcal Q = \mathds{CP}_{(1)}^{r- 1} \times \dots \times  \mathds{CP}_{(n)}^{r- 1} 
\subset  \mathds{CP}^{r^n - 1}, 
\end{eqnarray}
given by a product measure of the form 
\begin{eqnarray}
{\mathbb P}(\rd x) = {\mathbb P_1}(\rd x_1)\, {\mathbb P_2}(\rd x_2) \dots {\mathbb P_n}(\rd x_n),
\label{n particle measure}
 \end{eqnarray}
where the measure on each factor of the product space is assumed to be of the same type. 
Then it follows that the $H_j(x)$, $j = 1, 2, \dots, n$, when interpreted as functions on the bath state space, are independent identically distributed random variables under $\mathbb P$. 
As a consequence we see that the total Hamiltonian of the bath is given by a sum of $n$ independent identically distributed random variables. 
With this fact in mind, we can abbreviate the notation and omit the arguments of the functions in (\ref{bath Hamiltonian}), writing 
\begin{eqnarray}
H_B^{(n)}= \sum_{j = 1}^n H_j
\end{eqnarray}
for the total Hamiltonian of an $n$-particle bath. 

To develop a theory of the thermodynamics of such a system we shall take as our starting point a definition of the specific entropy associated with a given value $E$ of the specific energy.  In fact, we find it convenient to define $S(E)$ to be the specific entropy of the region of the state space for which the specific energy of the bath is no greater than $E$. In particular, in the case of $n$ particles we write
\begin{eqnarray}
S^{(n)}(E) =\frac {1}{n} \,  k_B \log \mathbb P\left[ \frac {1}{n} \sum_{j= 1}^{n} H_j \leq E\right].
\label{n-particle bath entropy}
\end{eqnarray}
Now, one might think that it would be better to define the specific entropy by confining the range of energy values to a thin band including $E$, say a closed set $[E - \Delta, E]$ for some $\Delta > 0$. One could proceed in that way, with some such band; but this is unnecessarily complicated since, as we show in Proposition  \ref{energy shell prop}, the specific entropy depends in the thermodynamic limit only on the upper boundary of the band. For large $n$, and specific energy $E$, ``most" of the entropy is concentrated in a thin shell just below $E$. As a consequence, we can stick with (\ref{n-particle bath entropy}), without the need for introducing a band. 

\section{Thermodynamic Limit}
\label{Thermodynamic Limit}

\noindent
Our strategy will be to show that for fixed $E$ the sequence $S^{(n)}(E)$, 
$n \in {\mathds N}$, converges for large $n$. The resulting expression 
\begin{eqnarray}
S(E) = \lim_{n \to \infty} S^{(n)}(E) 
\label{thermodynamic limit}
\end{eqnarray}
for the specific entropy  of the bath in the thermodynamic limit can then be used to work out the temperature of the bath, which is given as a function of the specific energy by 
 \begin{eqnarray}
 \frac { {\rm d} S(E)}  { {\rm d} E} = \frac {1} {T(E)}.
\label{temperature2}
\end{eqnarray}

To show that $S^{(n)}(E)$ converges we use a variant of Cram\'er's method in the theory of large deviations \cite{Cramer,Varadhan1,Ellis}. The result will be summarized in Proposition \ref{fundamental thermodynamic relation}. Our approach is to present a self-contained derivation of the thermodynamic limit, introducing the necessary mathematical tools as we go along, avoiding superfluous notions. As far as we are aware, we give here
the first general derivation of the thermodynamics of a system of weakly-interacting finite dimensional quantum systems. In particular, we do not make any assumptions concerning the choice to measure on the state spaces of the individual constituents, apart from a requirement of non-degeneracy, which we call the completeness condition. 

To begin,  we recall Markoff's inequality, which says that  if $a > 0$ is a constant and if $Y$ is a nonnegative random variable on a probability space 
$(\Omega, \mathcal F, \mathbb P) $, then 
 \begin{eqnarray}
\mathbb P(Y\geq a) \leq \frac{1}{a}\, \mathbb E[Y]. 
\label{Markov inequality}
\end{eqnarray}
Markov's inequality follows from the fact that if $Y \geq 0$ and $a > 0$, then 
$a \mathds{1}\{Y \geq a\} \leq Y$. Taking the expectation of each side, we obtain (\ref{Markov inequality}). 

Now suppose that $b \in {\mathds R}$ is a constant and $X$ is a random variable (not necessarily positive) such that 
$\mathbb E [\exp (-\theta X)] < \infty$ for all $\theta \geq 0$. 
Then it holds that
$\mathbb P(X\leq b) = \mathbb P(-\theta X \geq - \theta b) 
=  \mathbb P({\re}^ {- \theta X} \geq {\re} ^{- \theta b} )$
and thus by the Markov inequality we have 
\begin{eqnarray}
\mathbb P(X\leq b)  \leq  {\re} ^{ \theta b} \, {\mathbb E} \! \left[{\re}^ {- \theta X} \right] 
\label{pre-Chernoff} 
\end{eqnarray}
for all $\theta \geq 0$. Next we recall that if $c$ is a constant and $f(\theta)$ is a function of $\theta \geq 0$ such that $f(\theta) \geq c$ for all $\theta \geq 0$, then 
$\inf_{\theta \geq 0}f(\theta) \geq c$, where $\inf$ denotes the greatest lower bound. This is the ``tightness" property of the infimum. Thus optimizing (\ref{pre-Chernoff}) with respect to $\theta$ to obtain the tightest possible inequality we deduce that  
\begin{eqnarray}
\mathbb P(X\leq b) 
\leq  \inf_{\theta \geq 0} \, \re^{\theta b} \, {\mathbb E} \! \left[\re^ {-\theta X} \right] ,
\label{Chernoff inequality}
\end{eqnarray}
the so-called Chernoff bound. 
Applying this line of reasoning to the case of an $n$-particle bath one sees that by 
(\ref{pre-Chernoff}) we have 
\begin{eqnarray}
{\mathbb P} \left[ \frac {1} {n} \sum_{j= 1}^{n} H_j \leq E\right]  
\leq \re^{n \beta E} \, {\mathbb E} \! \left[ \exp \left ( - \beta \sum_{j= 1}^{n} H_j \right) 
\right]  
\end{eqnarray}
for all $n \in {\mathds N}$ and all $\beta  \in {\mathds R}^+$. Thus, writing 
\begin{eqnarray}
Z(\beta) = {\mathbb E} \left[ \exp\left( - \beta H \right) \right]
\end{eqnarray}
for the partition function of a representative molecule, and using the fact that the 
$\{H_j\}$ are independent identically distributed random variables, we obtain 
\begin{eqnarray}
\mathbb P\left[ \frac {1} {n} \sum_{j= 1}^{n} H_j \leq E\right]  
\leq  {\rm e}^{n \beta E} \, [Z(\beta)]^n,
\end{eqnarray}
and hence 
 \begin{eqnarray}
\frac {1} {n} \log \mathbb P\left[ \frac {1} {n} \sum_{j= 1}^{n} H_j \leq E\right]  
\leq  \beta E + \log Z(\beta) ,
\end{eqnarray}
which holds for all $n \in {\mathds N}$ and all $\beta  \in {\mathds R}^+$.  This leads us to the following tight bound, valid for all $n \in {\mathds N}$:
\begin{eqnarray}
\frac {1} {n} \log \mathbb P\left[ \frac {1} {n} \sum_{j= 1}^{n} H_j \leq E\right]  
\, \leq  \, \inf_{\beta \geq 0} \, \left [\beta E + \log Z(\beta)\right ].
\end{eqnarray}

Next, recall that for any sequence of real numbers $a_n$, $n \in {\mathds N}$, the superior limit is defined 
by $\limsup_{n \to \infty}  a_n = \lim_{n \to \infty} \sup_{m \geq n} a_m$ and the inferior limit is defined by $\liminf_{n \to \infty}  a_n = \lim_{n \to \infty} \inf_{m \geq n} a_m$. The superior limit and the inferior limit take values on the extended real line (including $\pm \infty$). In general the superior limit and the inferior limit need not be the same, but if the superior limit and the inferior limit agree, then their common value is defined to be the limit of the sequence. One can show that the superior limit has the property that if $b$ is a constant and if $a_n \leq b$ for all $n$, then 
$\limsup_{n \to \infty}  a_n \leq b$. With these facts in mind we deduce that
\begin{eqnarray}
\limsup_{n \to \infty} \frac {1} {n} \log \mathbb P\left[ \frac {1} {n} \sum_{j= 1}^{n} H_j \leq E\right]  
\, \leq  \, \inf_{\beta \geq 0} \, \left [\beta E + \log Z(\beta)\right ].
\label{upper bound} 
\end{eqnarray}

\section{Completeness condition}
\label{Completeness condition}
\noindent
To proceed, let us examine in more detail the expression appearing on the right side of this inequality. 
Write $E_-$ and $E_+$ respectively for the lowest and highest eigenvalues of the Hamiltonian $\hat H$ of a typical bath molecule, and write 
${\bar E}={\mathbb E}[H]$ for the mean under ${\mathbb P}$ of the associated random variable $H$. 

\begin{definition} 
We say that the measure ${\mathbb P}$ is $H$-complete if for any $\epsilon>E_-$ it 
holds that ${\mathbb P}(H<\epsilon)>0$.
\label{H complete}
\end{definition} 

\noindent Then we have the following. 

\begin{prop} 
If ${\mathbb P}$ is $H$-complete, it holds that
\begin{eqnarray}
\lim_{\beta\to\infty} \frac{{\mathbb E}\!\left[H \re^{-\beta H}\right]}
{{\mathbb E}\left[\re^{-\beta H}\right]} = E_- \,  ,   
\label{zero temperature limit}
\end{eqnarray} 
and for any $E\in(E_-,{\bar E}\,]$ there exists a unique value of $\beta\geq0$ such that 
\begin{eqnarray}
E = \frac{{\mathbb E}\!\left[H \re^{-\beta H}\right]}{{\mathbb E}\left[\re^{-\beta H}\right]} . 
\label{energy expectation} 
\end{eqnarray}
\label{completeness proposition}
\end{prop} 

\noindent {\bf Proof}.  
It should be apparent that the function defined by the right side of equation 
(\ref{energy expectation}) is continuous and decreasing with respect to $\beta$ for 
all $\beta\geq0$, for we have 
\begin{eqnarray}
\frac{\rd E}{\rd\beta} = - \frac{{\mathbb E}\!\left[(H-E)^2 \, \re^{-\beta H}\right]}
{{\mathbb E}\left[\re^{-\beta H}\right]} < 0, 
\end{eqnarray} 
and that at $\beta=0$ it takes the value ${\bar E}={\mathbb E}[H]$. Therefore, to 
establish the proposition it suffices to check that (\ref{zero temperature limit}) holds.
Equivalently, it we set $Y=H-E_-$ then $Y\geq0$ and it suffices to show that 
\begin{eqnarray}
\lim_{\beta\to\infty} \frac{{\mathbb E}\!\left[Y \re^{-\beta Y}\right]}
{{\mathbb E}\left[\re^{-\beta Y}\right]} = 0 .  
\label{Y inequality}
\end{eqnarray}
By Definition \ref{H complete}, ${\mathbb P}$ is $H$-complete if for any $\epsilon>0$ 
it holds that ${\mathbb P}(Y<\epsilon)>0$. If we let $\epsilon$ be given such that 
$\epsilon>0$, then we have 
\begin{eqnarray}
\frac{{\mathbb E}\!\left[Y \re^{-\beta Y}\right]}{{\mathbb E}\left[\re^{-\beta Y}\right]} = 
\frac{{\mathbb E}\!\left[{\mathds 1}\{Y<\epsilon\}Y \re^{-\beta Y}\right] + 
{\mathbb E}\!\left[{\mathds 1}\{Y\geq\epsilon\}Y \re^{-\beta Y}\right]}
{{\mathbb E}\left[{\mathds 1}\{Y<\epsilon\}\re^{-\beta Y}\right]+
{\mathbb E}\left[{\mathds 1}\{Y\geq\epsilon\}\re^{-\beta Y}\right]}.
\end{eqnarray}
But 
\begin{eqnarray}
{\mathbb E}\!\left[{\mathds 1}\{Y<\epsilon\}Y \re^{-\beta Y}\right]<\epsilon \, 
{\mathbb E}\!\left[{\mathds 1}\{Y<\epsilon\} \re^{-\beta Y}\right]
\end{eqnarray}
and 
\begin{eqnarray}
{\mathbb E}\!\left[{\mathds 1}\{Y\geq\epsilon\}Y 
\re^{-\beta Y}\right] < {E_+}\, {\mathbb E}\!\left[{\mathds 1}\{Y\geq\epsilon\}\re^{-\beta Y}\right].
\end{eqnarray}
It follows  that 
\begin{eqnarray}
\frac{{\mathbb E}\!\left[Y \re^{-\beta Y}\right]}{{\mathbb E}\left[\re^{-\beta Y}\right]} <
\frac{ \epsilon \, {\mathbb E}\left[{\mathds 1}\{Y<\epsilon\} \re^{-\beta Y}\right] + {E_+}\, 
{\mathbb E}\left[{\mathds 1}\{Y\geq\epsilon\} 
\re^{-\beta Y}\right] }{{\mathbb E}\left[{\mathds 1}\{Y<\epsilon\}\re^{-\beta Y}\right]+
{\mathbb E}\left[{\mathds 1}\{Y\geq\epsilon\}\re^{-\beta Y}\right]} .
\end{eqnarray}
Dividing both the denominator and the numerator by 
${\mathbb E}\!\left[{\mathds 1}\{Y<\epsilon\}\re^{-\theta Y}\right]$, 
which on account of the 
relation 
\begin{eqnarray}
{\mathbb E}\left[{\mathds 1}\{Y<\epsilon\}\re^{-\beta Y}\right]\geq \re^{-\beta\epsilon} \, 
{\mathbb E}\left[{\mathds 1}\{Y<\epsilon\}\right]=\re^{-\beta\epsilon}\, 
{\mathbb P}(Y<\epsilon) 
\end{eqnarray}
is strictly positive under the assumption that ${\mathbb P}$ is $X$-complete, we obtain
\begin{eqnarray}
\frac{{\mathbb E}\left[Y \re^{-\beta Y}\right]}{{\mathbb E}\left[\re^{-\beta Y}\right]} <
\frac{\epsilon + {E_+}\, R(\epsilon,\beta)} {1 + R(\epsilon,\beta)} , 
\end{eqnarray}
where 
\begin{eqnarray}
R(\epsilon,\beta) = \frac{{\mathbb E}\left[{\mathds 1}\{Y\geq\epsilon\} 
\re^{-\beta Y}\right] }{{\mathbb E}\left[{\mathds 1}\{Y<\epsilon\}\re^{-\beta Y}\right]}  . 
\end{eqnarray}
We shall show that $\lim_{\beta\to\infty}R(\epsilon,\beta)=0$ for any choice of $\epsilon>0$. 
Recall that if $f(x)$ is convex and if $Y$ and $f(Y)$ are integrable, then by Jensen's inequality we have ${\mathbb E}[f(Y)] \geq 
f({\mathbb E}[Y])$. More generally, suppose that $B$ is any nonnegative random variable such that
$0 <   {\mathbb E}[B] < \infty$.  Then if $f(x)$ is convex and if $BY$ and $Bf(Y)$ are integrable, it holds that 
\begin{eqnarray}
\frac { {\mathbb E}[B f(Y)] } { {\mathbb E}[B] } \geq f \left ( \frac { {\mathbb E}[B Y] } { {\mathbb E}[B] } \right ) .
\end{eqnarray}
Since $\re^{-\beta x}$ is convex, we see that 
\begin{eqnarray}
\frac{{\mathbb E}\left[{\mathds 1}\{Y<\epsilon\} 
\re^{-\beta Y}\right]} {{\mathbb E}[{\mathds 1}\{Y<\epsilon\}]} \geq 
\exp\left( -\beta\langle Y\rangle_{Y<\epsilon} \right) , 
\end{eqnarray}
where 
\begin{eqnarray}
\langle Y\rangle_{Y<\epsilon} = \frac{{\mathbb E}[{\mathds 1}\{Y<\epsilon\} 
Y] }{{\mathbb E}[{\mathds 1}\{Y<\epsilon\}]} . 
\end{eqnarray}
It follows that 
\begin{eqnarray}
R(\epsilon,\beta) \leq \frac{{\mathbb E}\left[{\mathds 1}\{Y\geq\epsilon\} 
\exp\big(-\beta (Y-\langle Y\rangle_{Y<\epsilon})\big)\right] }{{\mathbb E}[{\mathds 1}\{Y<\epsilon\}]}  . 
\end{eqnarray}
But $Y-\langle Y\rangle_{Y<\epsilon}>0$ for all $Y\geq\epsilon$. Therefore 
$\lim_{\beta\to\infty}R(\epsilon,\beta)=0$, and thus
\begin{eqnarray}
\lim_{\beta\to\infty} \frac{{\mathbb E}\left[Y \re^{-\beta Y}\right]}
{{\mathbb E}\left[\re^{-\beta Y}\right]} < \epsilon 
\end{eqnarray}
for any choice of $\epsilon>0$, which implies (\ref{Y inequality}) since $Y$ is nonnegative. 
\hfill $\Box$
\vspace{5.00pt} 

Returning to the expression on the right side of equation (\ref{upper bound}), we conclude that if ${\mathbb P}$ 
is $H$-complete then there exists a unique value of $\beta\geq0$ such that the infimum is obtained for 
any given value of $E$ in the range $(E_-,{\bar E}]$, namely, the value of $\beta$ such that equation 
(\ref{energy expectation}) is satisfied. For each value of $E\in(E_-,{\bar E}]$ let us write $\beta(E)$ for the 
corresponding value of $\beta$. Then we have 
\begin{eqnarray}
\inf_{\beta\geq0} \big[ \beta E + \log Z(\beta) \big] = \beta(E) \, E + \log Z(\beta(E)) . 
\end{eqnarray} 
Inserting this expression for the infimum back into (\ref{upper bound}) we thus obtain the inequality
\begin{eqnarray}
 \limsup_{n \to \infty} \frac {1} {n} \log \mathbb P\left[ \frac {1} {n} \sum_{j= 1}^{n} H_j \leq E\right]  
\, \leq  \, \beta(E) \, E +   \log Z(\beta(E)) .
\label{upper bound 2} 
\end{eqnarray}

\section{Law of Large Numbers}
\label{Law of Large Numbers}
\noindent
Going forward, we need the weak law of large numbers. Recall that if $\{Y_j\}$ for $j \in {\mathds N}$ is a sequence of random variables on a probability space $(\Omega, \mathcal F, \mathbb P)$ then we say that $Y_j$ 
converges in probability (under $\mathbb P$)  to the random variable $Y$ if for any 
$\delta > 0$ it holds that 
\begin{eqnarray}
 \lim_{j \to \infty}  \mathbb P\left[ \, \left | Y_j - Y  \right | > \delta \, \right ]  = 0. 
\label{convergence in probability} 
\end{eqnarray}
We recall also that if a random variable $X$ has finite mean $\mu$ and variance $\sigma^2$ then for any $\delta > 0$ we have the Chebychev inequality
\begin{eqnarray}
{\mathbb P}\left[ \, \left| X - \mu \right| \geq \delta \, \right]  
 &=& {\mathbb E} \left[ \mathds{1} \{ ( X - \mu )^2 \geq \delta^2 \} \right] \nonumber \\
 &\leq&  \frac {1}{\delta^2} \, {\mathbb E}\left[ \mathds{1} \{ ( X - \mu )^2 \geq \delta^2 \} 
 (X - \mu )^2 \right] \nonumber \\
 &\leq&  \frac{1}{\delta^2} \, {\mathbb E} \left[ (X - \mu )^2 \right] \nonumber \\
 &=&  \frac{1}{\delta^2} \, \sigma^2.
\label{convergence} 
\end{eqnarray}
Let $\{X_j\}_{j \in \mathds N}$ be a sequence of iid random variables with mean 
$\mu$ and variance $\sigma^2$. Then by the Chebychev inequality and the iid 
property we obtain 
\begin{eqnarray}
{\mathbb P}\left[\, \left | \, \frac {1}{n} \sum_{j=1}^nX_j - \mu \, \right | \,  \geq \delta \right ]  
\leq \frac {1} {n\delta^2} \, \sigma^2 
\label{Chebychev} 
 \end{eqnarray}
for all $n \in \mathds{N}$, and hence it follows that for all $\delta > 0$ we have 
\begin{eqnarray}
\lim_{n \to \infty} \, \mathbb P\left[\, -\delta < \frac {1}{n} \sum_{j=1}^nX_j - \mu  < \delta \right ]  
= 1, 
 \end{eqnarray}
the weak law of large numbers.
Going forward  we require the technique of ``change of measure". On a probability space $(\Omega, \mathcal F, \mathbb P)$ let the random variable $Z$ be such that $Z>0$ and ${\mathbb E}[Z] < \infty$, and for each measurable set $A \in \mathcal F$ write  
\begin{eqnarray}
{\mathbb Q}(A) = \frac{{\mathbb E}[Z\, {\mathds 1}\{A\}]}{{\mathbb E}[Z]}. 
\end{eqnarray}
Then $\mathbb Q$ defines a new probability measure on the measurable space $(\Omega, \mathcal F)$, and we refer to the transformation $(\Omega, \mathcal F, \mathbb P) \to (\Omega, \mathcal F, \mathbb Q)$ as a change of measure. If $Z$ takes the form 
$Z = { \rm e}^ { -\theta Y}$ for some random variable $Y$ such that $\mathbb E [ { \rm e}^ { -\theta Y} ] < \infty$
for a nontrivial range of values of $\theta$ containing the origin, then for each such value of $\theta$ in the resulting family of transformation we refer to the measure change $\mathbb P \to \mathbb P^{\theta}$ as an Esscher transformation \cite{Esscher}.  

In the context of a quantum heat bath consisting of $n$ molecules, recall that if $H$ denotes the Hamiltonian function associated with a representative molecule then $\mathbb E [H] = \bar E$, the mean energy under $\mathbb P$. If for a fixed value of the specific energy $E$ of the bath we define the corresponding inverse temperature by $\beta(E)$, then under the Esscher transformation 
$\mathbb P \to \mathbb P^{\beta}$ induced by the factor $Z = { \rm e}^ { -\beta H}$ associated with  $H$  we have $\mathbb E^{\beta} [H] = E$. This follows from the fact that 
\begin{eqnarray}
{\mathbb E}^{\beta} \left[H \right] = \frac { {\mathbb E} \left[{ \rm e}^ { -\beta H} H \right] } {Z(\beta)} .
\end{eqnarray}
More generally, for each value of $n$ it holds that 
\begin{eqnarray}
{\mathbb E}^{\beta} \left[ \frac{1}{n} \sum_{j= 1}^{n} H_j \right] = E.  
\end{eqnarray}
Here we extend the Esscher transformation to include a factor for each molecule by setting
\begin{eqnarray}
{\mathbb E}^{\beta} \left[ Y \right] = \frac {1}{(Z(\beta))^n} \mathbb E\left[\, \exp \left(-\beta \sum_{j= 1}^{n} H_j  \right) Y
 \right]  
\end{eqnarray}
for any integrable random variable $Y$. As a consequence, the weak law of large numbers takes a different form under $\mathbb P^{\beta}$ from what it does under $\mathbb P$. In particular, whereas under $\mathbb P$ the weak law of large numbers  says that for all 
$\delta > 0$ we have
\begin{eqnarray}
\lim_{n \to \infty} \, \mathbb P\left[\, -\delta < \frac {1}{n} \sum_{j=1}^n H_j - \bar E  < \delta \right] = 1 ,
\label{weak law E bar} 
\end{eqnarray}
we find that under  $\mathbb P^{\beta}$ the weak law of large numbers takes the form
\begin{eqnarray}
\lim_{n \to \infty} \, \mathbb P^{\beta} \left[\, -\delta < \frac {1}{n} \sum_{j=1}^n H_j - E  < \delta \right]  = 1 ,
\label{weak law for heat bath}
\end{eqnarray}
in which the ${\bar E}$ in (\ref{weak law E bar}) is replaced with an $E$ in (\ref{weak law for heat bath}). 

With these preliminaries at hand, we are in a position to establish a further inequality relevant to the calculation of the specific entropy of a quantum heat bath in the thermodynamic limit. 
Let $\epsilon > 0$ be given, and choose $\delta$ so that $0 < \delta < \epsilon$. Then for any fixed value of $E$ we have
 \begin{eqnarray}
 && \!\!
\mathbb P\left[ \frac {1} {n} \sum_{j= 1}^{n} H_j < E + \epsilon \right] 
= \mathbb E\left[\,\mathds {1} \left\{ \frac {1} {n} \sum_{j= 1}^{n} H_j < E 
+ \epsilon \right \} \right]  \nonumber \\ && \quad
\geq \mathbb E\left[\,\mathds {1} \left\{  E - \delta < \frac {1} {n} 
\sum_{j= 1}^{n} H_j < E + \delta \right \} \right] \nonumber \\
&& \quad
\geq \mathbb E\left[\, \exp \left[-\beta \left(\sum_{j= 1}^{n} H_j  -n(E - \delta) \right) \right]
\mathds {1} \left\{  E - \delta < \frac {1} {n} \sum_{j= 1}^{n} H_j < E + \delta \right \} \right]. 
\end{eqnarray}
It follows by a change of measure in the expectation in the line above that
 \begin{eqnarray}
&& \!\! 
{\mathbb P}\left[ \frac {1} {n} \sum_{j= 1}^{n} H_j < E + \epsilon \right] 
 \nonumber \\ && \quad 
 \geq \exp \left[n \beta (E - \delta) \right] 
[Z(\beta)]^n \,
\mathbb E^{\beta} \left[\, 
\mathds {1} \left\{  E - \delta < \frac {1} {n} \sum_{j= 1}^{n} H_j < E + \delta \right \} \right] ,
\end{eqnarray}
and therefore
 \begin{eqnarray}
 && \! \! 
\frac {1} {n} \log \mathbb P\left[ \frac {1} {n} \sum_{j= 1}^{n} H_j < E + \epsilon \right] 
 \nonumber \\ && \quad \geq 
 \beta (E - \delta) + 
\log Z(\beta) \, +  \frac {1} {n} \log
\mathbb E^{\beta} \left[\, 
\mathds {1} \left\{  E - \delta < \frac {1} {n} \sum_{j= 1}^{n} H_j < E + \delta \right \} \right] .
\end{eqnarray}
Then taking the inferior limit of each side of this inequality, and using the weak law of large numbers in the form (\ref{weak law for heat bath}) we obtain
 \begin{eqnarray}
\liminf_{n \to \infty} \frac {1} {n} \log \mathbb P\left[ \frac {1} {n} \sum_{j= 1}^{n} H_j < E + \epsilon \right] 
\geq  \beta (E - \delta) + \log Z(\beta) .
\label{with delta and epsilon}
\end{eqnarray}
Since $\epsilon >0$ is arbitrary and (\ref{with delta and epsilon})  holds for any $\delta >0$ such that $\delta < \epsilon$, 
it must hold for $\delta$ and $\epsilon$ arbitrarily small, and we conclude that 
 \begin{eqnarray}
\liminf_{n \to \infty} \frac {1} {n} \log \mathbb P\left[ \frac {1} {n} \sum_{j= 1}^{n} H_j \leq E \right] 
\geq  \beta(E) \,E + \log Z(\beta(E)) ,
\label{lower bound}
\end{eqnarray}
where we have restored the dependence of $\beta$ on $E$ to emphasize that (\ref{lower bound}) holds for the value of $\beta$ determined by equation (\ref{energy expectation}). 

In more detail, to obtain (\ref{lower bound}) we observe that, for fixed $\epsilon$,  (\ref{with delta and epsilon}) tells us that for all $\delta > 0$ an inequality of the form $ y \geq z - \beta \delta$ holds, where one can read off the expressions for $y$ and $z$. Now, if $ y \geq z - \beta \delta$ for all $\delta > 0$, then $ y\geq z$. For suppose $ y < z$. Then there exists a $q$ such that 
$ y < q < z$, and hence $ y < z - (z - q)$. But that implies $ y < z - \beta \delta_0$ with $\delta_0 = (z - q)/\beta >0$ which contradicts our assumption that $ y \geq z - \beta \delta$ for $\delta > 0$. Hence $ y\geq z$, and we conclude that (\ref{with delta and epsilon})  holds with $\delta = 0$ for all $\epsilon > 0$. Next we observe that if
 \begin{eqnarray}
\liminf_{n \to \infty} \frac {1} {n} \log \mathbb P\left[ \frac {1} {n} \sum_{j= 1}^{n} H_j < E + \epsilon \right] 
\geq  \beta E + \log Z(\beta) 
\end{eqnarray}
for all $\epsilon > 0$, then by the tightness property of the infimum we have
 \begin{eqnarray}
\inf_{\epsilon > 0} \,  \liminf_{n \to \infty} \frac {1} {n} \log \mathbb P\left[ \frac {1} {n} \sum_{j= 1}^{n} H_j < E + \epsilon \right] 
\geq  \beta E + \log Z(\beta) .
\end{eqnarray}
However,
 \begin{eqnarray}
\inf_{\epsilon > 0} \,  \liminf_{n \to \infty} \frac {1} {n} \log \mathbb P\left[ \frac {1} {n} \sum_{j= 1}^{n} H_j < E + \epsilon \right] 
= \liminf_{n \to \infty} \frac {1} {n} \log \mathbb P\left[ \frac {1} {n} \sum_{j= 1}^{n} H_j \leq E  \right] ,
\end{eqnarray}
and that leads to (\ref{lower bound}), as claimed.

Now we are in a position to derive the fundamental thermodynamic relation
(\ref{entropy formula}).
For if we compare the inequalities (\ref{upper bound 2}) and (\ref{lower bound}) and use the fact that for any sequence
of real numbers $\{a_n\}_{n \in \mathds N}$ it holds that $\limsup_{n \to \infty} a_n \geq \liminf_{n \to \infty} a_n$, we immediately conclude that the expression on the left side of (\ref{upper bound 2}) and the expression on the left side of (\ref{lower bound}) must be equal. 
Thus, we have the following. 

\begin{prop}
\label{fundamental thermodynamic relation}
The thermodynamic limit
 \begin{eqnarray}
S(E) =  \lim_{n \to \infty} \frac {1}{n}\, k_B 
\log \mathbb P\left[ \frac {1}{n} \sum_{j= 1}^{n} H_j \leq E\right] 
\label{thermodynamic limit 2}
\end{eqnarray}
exists, and the resulting expression for the specific entropy of the heat bath is 
 \begin{eqnarray}
S(E) =  k_B \,  \beta(E) \, E + k_B \, \log Z(\beta(E)),
\label{specific entropy 2}
\end{eqnarray}
where for each value of $E \in (E_-, \bar E ]$ the associated value of $\beta$ is determined by
\begin{eqnarray}
E = \frac{{\mathbb E}\!\left[H \re^{-\beta(E) H}\right]}{{\mathbb E}\left[\re^{-\beta(E) H}\right]} , 
\end{eqnarray}
and
$Z(\beta(E)) = {\mathbb E} \left[ \exp\left( - \beta(E) H \right) \right]$.

\end{prop}

\section{Independence of width of energy shell}
\label{Independence of width of energy shell}
\noindent
It is interesting to observe, as we remarked earlier, that value of the entropy of a quantum heat bath in the thermodynamic limit is insensitive to the width of the band of energies below the specific energy. More precisely, let us write
 \begin{eqnarray}
S^{(n)}(E- \Delta, E) =\frac {1}{n} \,  k_B \log \mathbb 
P\left[ E- \Delta \leq \frac {1}{n} \sum_{j= 1}^{n} H_j \leq E\right] 
\label{n-particle entropy with energy band}
\end{eqnarray}
for the specific entropy of an $n$-particle system of weakly interacting particles when the 
specific energy of the system lies in the band $[E- \Delta, E]$ for some $\Delta >0$. We do 
not require that $\Delta$ should be small, though one might have that case in mind. We 
shall show that the thermodynamic limit of (\ref{n-particle entropy with energy band}) exists, 
and that the resulting expression for the specific entropy of the shell is independent of $\Delta$. 
We have the following. 
\begin{prop}
\label{energy shell prop}
The limit 
$S(E- \Delta, E)  = \lim_{n \to \infty} S^{(n)}(E- \Delta, E)$
exists for the entropy associated with an energy shell, and is given by 
$S(E- \Delta, E) = S(E)$.
\end{prop}
Here $S(E)$ is  the expression given by (\ref{specific entropy 2}), obtained without the specification of the lower bound of the energy shell. This result may seem surprising at first glance, but one is familiar with many instances of calculations in statistical mechanics where the device of a band of energies is introduced, only for the relevant physical results later not to depend on it. The interpretation of the situation in the present context is that in the thermodynamic limit the specific entropy, for a given specific energy, is concentrated almost entirely in a thin shell of the quantum phase space at and immediately below the energy surface corresponding to the given value of the specific energy. 

The result can be understood as an example of the idea of ``concentration of measure", but is 
perhaps more easily understood in probabilistic terms. For the given \textit{a priori} measure it 
is extremely unlikely that the average of the energies of a large number of independent molecules 
will be anything other than the \textit{a priori} mean $\bar E$, but if we condition on the average 
being no greater than some specified value $E$, then it will be extremely unlikely that the average 
will be much less than $E$.  

Putting the matter differently, we remark that for large $n$ the least unlikely of all the unlikely events will necessarily dominate.
Even in the case of the Dirac measure, which corresponds to the situation usually considered in quantum statistical mechanics based on the enumeration of energy eigenstates, the effect of this concentration of measure to a very thin band ensures that the dependence of the entropy on the choice of the width $\Delta$ drops out in the thermodynamic limit, and that the temperature, which is ill defined for any finite $n$ under the Dirac measure, is well defined in the thermodynamic limit.
These conclusions are consistent with the results obtained in \cite{griffiths}. 
\vspace{6.00pt} 

\noindent {\bf Proof of Proposition \ref{energy shell prop}}.  
As a consequence of (\ref{upper bound 2}) it holds for any choice of 
$\Delta > 0$ that
\begin{eqnarray}
 \limsup_{n \to \infty} \frac {1} {n} \log \mathbb 
 P\left[ E- \Delta \leq \frac {1} {n} \sum_{j= 1}^{n} H_j \leq E\right]  
\, \leq  \, \beta(E) \, E +   \log Z(\beta(E)) .
\label{upper bound 3}
\end{eqnarray}
Now let $\Delta > 0$ and $\epsilon > 0$ be given, and choose $\delta$ so that $0 < \delta <  \min(\Delta, \epsilon)$. We see that
 \begin{eqnarray}
\mathbb P\left[ E- \Delta \leq \frac {1} {n} \sum_{j= 1}^{n} H_j < E + \epsilon \right] 
= \mathbb E\left[\,\mathds {1} \left\{ E- \Delta \leq \frac {1} {n} \sum_{j= 1}^{n} H_j < E + \epsilon \right \} \right] 
\nonumber \\ 
\geq \mathbb E\left[\,\mathds {1} \left\{  E - \delta \leq \frac {1} {n} \sum_{j= 1}^{n} H_j \leq E + \delta \right \} \right] .
\end{eqnarray}
From this point forward, the argument proceeds as in the line of reasoning involving a change of measure and the weak law of large numbers leading to (\ref{lower bound}), only now we obtain
 \begin{eqnarray}
\liminf_{n \to \infty} \frac {1} {n} \log \mathbb P\left[ E- \Delta \leq \frac {1} {n} \sum_{j= 1}^{n} H_j \leq E \right] 
\geq  \beta(E) \,E + \log Z(\beta(E)) .
\label{lower bound 2}
\end{eqnarray}
Comparing (\ref{upper bound 3}) and (\ref{lower bound 2}), and using  the fact that the superior limit dominates the inferior limit, we deduce that  
 \begin{eqnarray}
\lim_{n \to \infty} \frac {1} {n} \log \mathbb P\left[ E- \Delta \leq \frac {1} {n} \sum_{j= 1}^{n} H_j \leq E \right] 
= \beta(E) \,E + \log Z(\beta(E)) ,
\end{eqnarray}
and hence
 \begin{eqnarray}
S(E- \Delta, E) =  k_B \, \beta(E) \, E + k_B \, \log Z(\beta(E)) . 
\end{eqnarray}
This shows that in the thermodynamic limit the specific entropy of the heat bath is independent of the lower bound of the energy shell.  
\hfill $\Box$
%
\section{Examples}
\label{examples}
\noindent
To gain further intuition about the thermodynamics
of a quantum heat bath, it will be instructive if we examine some specific examples. We begin with the 
Dirac measure. This is the case when the measure $\mathbb{P}$ on the state space of a representative molecule is concentrated on the energy eigenstates. As before, the Hilbert space associated with an individual molecule is taken to have dimension $r$. In the situation that the Hamiltonian has a nondegenerate spectrum the Dirac measure is given by
\begin{eqnarray}
{\mathbb P}(\rd x) = \frac{1}{r} \sum_i \delta_{i}(\rd x ). 
\label{Boltzmann measure} 
\end{eqnarray}
Here $\delta_{i}(\rd x )$ denotes the usual Dirac measure concentrated at the point $x_i$
$(i = 1, 2, \dots, r)$, where $x_i$  denotes for each  $i$ the point in the state apace 
$\mathit \Gamma$ corresponding to the energy 
eigenstate $|x_i\rangle$ with energy $E_i$. In the case of a Hamiltonian with a degenerate spectrum, the situation is a little more complicated, but one expects this. In that case we understand $\delta_{i}(\rd x )$ to represent for each value of $(i = 1, 2, \dots, r)$ the uniform measure concentrated (with total mass unity) on the projective subspace consisting of states of energy $E_i$. If the multiplicity of level $i$ is $m_i$, then $\delta_{i}(\rd x )$ is uniform on the $(m_i - 1)$-dimensional projective space consisting of states of energy $E_i$, and vanishes elsewhere. Consider, for instance, a Hilbert space of dimension three. In the nondegenerate case, the measure is concentrated at the eigenstates $x_1$, $x_2$, and $x_3$, corresponding to the energy levels $E_1$, $E_2$, and $E_3$. As an example of the degenerate case, suppose, say, that $E_2 = E_3$, and that $E_1$ is distinct. If $x_2$ and $x_3$ have the same energy, then any superposition of these states is also an eigenstate with that energy. Thus we obtain a complex projective line of eigenstates. This line is given the uniform measure, and is counted twice according to (\ref{Boltzmann measure}). Thus the Dirac measure in this case consists of the usual Dirac measure, with overall weight 1/3, at  
$x_1$, together with a uniform measure concentrated on the line joining $x_2$ and $x_3$, with overall weight 2/3. 

In the case of the Dirac measure, it should be apparent that for $r = 2$ the partition function 
(\ref{partition function}) is 
\begin{eqnarray}
Z(\beta) = \half \left( \re^{-\beta E_1} +  \re^{-\beta E_2} \right) ,
\label{Boltzmann partition function} 
\end{eqnarray}
which, apart from the factor of $\half$, is the well known formula one finds in standard textbook 
treatments of quantum statistical mechanics. 
In other words, if ${\mathbb P}$ assigns probability $\frac{1}{2}$ to each of 
the energy eigenstates, and probability zero to all other states, then the Esscher transformed measure 
${\mathbb P}^\beta$ assigns the usual Boltzmann weights 
$p_1 = \re^{-\beta E_1}/ ( \re^{-\beta E_1} +  \re^{-\beta E_2} )$ and 
$p_2 = \re^{-\beta E_2}/( \re^{-\beta E_1} +  \re^{-\beta E_2} )$ to the energy eigenstates, and we are led back to the standard treatment. 
It should be noted, however, that since ${\mathbb P}^\beta$ is absolutely absolute continuous with ${\mathbb P}$, it follows that the Dirac measure is the only 
choice of $\mathbb P$ that gives rise to the expression (\ref{Boltzmann partition function}) for the partition function: if nonzero probabilities 
are assigned to superpositions of energy eigenstates, then the resulting partition function will take a 
form different from that of (\ref{Boltzmann partition function}). Thus, there is an element of incompatibility 
between the superposition principle of quantum mechanics, and the standard treatment of 
quantum statistical mechanics. 

Continuing the analysis, we find that the associated expression in this case for the specific energy (\ref{specific energy}), 
as a function of $\beta$, is
\begin{eqnarray}
E(\beta) = \frac {E_1 \, \re^{-\beta E_1} + E_2 \, \re^{-\beta E_2} }
{ \re^{-\beta E_1} +  \re^{-\beta E_2} } . 
\label{Boltzmann 2}
\end{eqnarray}
The mean energy $\bar E = {\mathbb E}[H(x)]$ under $\mathbb P$ is then given by 
\begin{eqnarray}
\bar E = \half (E_1 + E_2), 
\end{eqnarray}
and it is clear that for each value of $E \in (E_1, \bar E]$ there exists a value of 
$\beta \in {\mathds R}^+$ such that (\ref{Boltzmann 2}) is satisfied. In fact, we can invert this relation, to give $\beta$ as a function of $E$, as follows: 
\begin{eqnarray}
\beta(E) = \frac {1}{E_2 - E_1} \log \frac {E_2 - E} {E - E_1}.
\label{beta in terms of E}
\end{eqnarray}
Inserting this expression for $\beta$ in terms of $E$ back into the partition function, we obtain a formula for a partition function as a function of $E$, given by 
\begin{eqnarray}
Z(\beta(E)) = \frac{1}{2} \left[  \left( \frac  {E - E_1} {E_2 - E_1} \right)^{\frac {E_1}  {E_2 - E_1}}  +
 \left(  \frac {E_2 - E} {E_2 - E_1} \right)^{\frac {E_2}  {E_2 - E_1}}   \right].
\label{Z in terms of E}
\end{eqnarray}
Finally, inserting (\ref{beta in terms of E}) and (\ref{Z in terms of E}) into the thermodynamic relation 
(\ref{specific entropy 2}), we obtain the following expression for the specific entropy of the bath as a function of the specific energy, which is
valid for $E \in (E_1, \bar E]$: 
\begin{eqnarray}
S(E)= k_B    \left[  \log \half - \frac  {E - E_1} {E_2 - E_1}  \log \frac  {E - E_1} {E_2 - E_1} 
- \frac  {E_2 - E} {E_2 - E_1} \log \frac  {E_2 - E} {E_2 - E_1}  \right].  
\label{entropy in terms of E}           
\end{eqnarray}
To get a feeling for this formula, set 
\begin{eqnarray}
p =  \frac  {E - E_1} {E_2 - E_1} .     
\end{eqnarray}
Then (\ref{entropy in terms of E}) takes the form of a Shannon entropy associated with a binary distribution: 
\begin{eqnarray}
S(E)= k_B    \left[  \log \half - p \log p - (1-p) \log \, (1-p) \right].  
\label{information theory}        
\end{eqnarray}
The constant term involving $\log \half$ has the effect of ensuring that the maximum value of the entropy is zero, which occurs at $p = \half$, or equivalently at $E = \half (E_1 + E_2)$, that is to say, at infinite temperature. 

It can be useful for some purposes to separate out the infinite temperature limit in the 
familiar expression (\ref{Boltzmann 2}). When one does this, the terms left over depend on the difference between the two energy levels. More specifically, if we set $\omega = \half(E_2 - E_1)$, then 
\begin{eqnarray}
E(\beta) =\half (E_1 + E_2) - \omega \tanh (\beta \omega).
\label{Boltzmann 3}
\end{eqnarray}
We can use 
(\ref{Boltzmann 3}) together with (\ref{heat capacity}) to work out the heat capacity of the bath in the case of the Dirac measure, which is given by 
\begin{eqnarray}
C(\beta) = k_B \beta^2 \omega^2  \rm{sech}^2  (\beta \omega),
\label{heat capacity Boltzmann}
\end{eqnarray}
and one can check that this goes to zero at low temperature, as is well known. 
Then if we make use of the Taylor series expansion 
\begin{eqnarray}
\tanh x = x -  \tfrac {1}{3} x^3 + \tfrac{2}{15} x^5 + \cdots
\end{eqnarray}
we are led to an expansion for the energy as a function of the inverse temperature
\begin{eqnarray}
E(\beta) =\half (E_1 + E_2) - \beta \omega^2 + \tfrac {1}{3} \beta^3 \omega^4 + \cdots,
\label{Boltzmann 4}
\end{eqnarray}
which can be used for comparison with other models at high temperature. 

One should bear in mind that all the formulae above are to be interpreted as being applicable at the macroscopic level, that is to say, at the level of the thermodynamic properties of substance under consideration. Thus we see that the specification of the Hamiltonian of a representative molecule at the microscopic level, along with the specification of the relevant measure on the state space of the molecule (in this case, the Dirac measure), is sufficient to determine completely the equation of state of the bath, in the form of a relation between the energy and the entropy of the bath system as a whole. 

Now we turn to the
uniform measure, or Haar measure, which  in the case of the state space of a single molecule is given by an expression of the form
\begin{eqnarray}
{\mathbb P}(\rd x) = \frac{1}{V_{\mathit \Gamma} } \, \rd V_x  \, . 
\end{eqnarray}
Here $\rd V_x$ denotes the natural volume element associated with the Fubini-Study metric on $\mathit \Gamma$, and $V_{\mathit \Gamma}$ is the total volume of $\mathit \Gamma$. In the case of $n$ weakly interacting molecules the uniform measure is defined as in the product (\ref{n particle measure}), with a uniform measure on the phase space of each molecule. As we have suggested, the uniform measure is in some respects the most natural choice of a measure on the phase space of the bath, since it involves no specification of additional structure, apart from that already implicit in assuming the the bath molecules are mutually disentangled. In fact, we have already seen how the uniform measure arises in the previous example in the situation where there are degeneracies. 
 
The calculations that arise in connection with the Dirac measure leading to the Boltzmann weights are, generally speaking, familiar to physicists; but the uses of the uniform measure are less familiar. As an illustration of typical calculations involving the uniform measure on the Fubini-Study manifold, we work out the mean and the variance of the Hamiltonian function  of a bath molecule. Let us write 
$H(x)=H^a_b \Pi^b_a(x)$, using the summation convention, where $H^a_b$ denotes the 
matrix elements of the Hamiltonian ${\hat H}$ of the molecule, in a suitable basis, and 
$\Pi^b_a(x)$ denotes the matrix elements of 
the projection operator $|x\rangle \langle x|/\langle x|x\rangle$ corresponding to a point $x$ in the state space of the molecule. Then for the mean under the uniform measure we have
\begin{eqnarray}
{\bar E} = \int_{\mathit \Gamma} H(x) {\mathbb P}(\rd x) = H^a_b \int_{\mathit \Gamma} \Pi^b_a(x) 
 {\mathbb P}(\rd x) = \frac{1}{r} \, H^a_b \delta^b_a , 
 \label{uniform mean} 
\end{eqnarray} 
where $r$ is the dimension of the Hilbert space. In other words, it holds that
\begin{eqnarray}
{\bar E}= \frac {1} {r} \,{\rm tr} \, {\hat H},
\end{eqnarray} 
which is the same as the result obtained for the mean 
 under the Dirac measure. For the calculation of the uniform average of the 
projection operator, see, e.g., \cite{GWG}. 

On the other hand, the infinite-temperature statistics associated 
with the Dirac measure and the uniform measure disagree at the second moment. Under the Dirac measure 
we have 
\begin{eqnarray}
{\mathbb E}[H^2(x)] = \frac {1} {r} \, {\rm tr}({\hat H}^2),
\end{eqnarray} 
leading to a variance of
\begin{eqnarray}
\sigma^2 = \overline{E^2} - {\bar E}^2 ,
\end{eqnarray} 
where for the average of the squares of the energy eigenvalues we have written 
\begin{eqnarray}
{\overline {E^2}}= \frac {1} {r} \,{\rm tr} \, {\hat H^2}.
\end{eqnarray} 
Under the uniform measure, however, 
we have 
\begin{eqnarray}
{\mathbb E}[H^2(x)]  &=& \int_{\mathit \Gamma} H^2(x) \, {\mathbb P} (\rd x)  \nonumber \\ &=& 
H^a_b H^c_d \int_{\mathit \Gamma} \Pi^b_a(x) \Pi^d_c(x) {\mathbb P}(\rd x) \nonumber \\ &=& 
\frac{1}{r(r+1)}  H^a_b H^c_d (\delta^b_a \delta^d_c + \delta^d_a\delta^b_d) 
\nonumber \\ &=& 
\frac{1}{r(r+1)} \left( {\rm tr}({\hat H}^2) + ({\rm tr}({\hat H}))^2 \right),  
 \label{variance calculation} 
\end{eqnarray}
leading to a variance of 
\begin{eqnarray}
\sigma^2 = \frac{1}{r+1}(\overline{E^2} - {\bar E}^2 ).
\end{eqnarray} 
For instance  the case of a two-dimensional Hilbert space with energy levels $E_1$ and $E_2$ it holds that $\sigma^2 = \frac{1}{4}(E_2-E_1)^2$ under the Dirac measure, whereas $\sigma^2 = \frac{1}{12}(E_2-E_1)^2$ under the uniform measure. 

In the case $r = 2$ we find, more generally, that the Hamiltonian function $H(x)$, when viewed as a random variable, has a uniform distribution over the interval $[E_1, E_2]$. That is, 
\begin{eqnarray}
\mathbb{P} (H \leq E) = \mathds 1 \{ E_1 \leq E \leq E_2 \} \frac {E - E_1} {E_2 - E_1} + \mathds 1 \{ E_2<E \}.
\end{eqnarray}
The proof of this fact is as follows. It is well known that on a complex projective space of one dimension the Fubini-Study metric 
is equivalent to that of the ordinary two-sphere. Thus if we introduce polar coordinates $\theta, \phi$ such that 
$0 \leq \theta \leq \pi$ and $0 \leq \phi < 2\pi$ then
\begin{eqnarray}
\mathbb{P} (\rd x) = \frac {1} {4\pi} \sin \theta \, \rd \theta \, \rd \phi,
\end{eqnarray}
where the factor of $4\pi$ in the denominator ensures that the total measure of the surface of the sphere is normalized to unity. In this way the Fubini-Study manifold is given the structure of a probability space, and we can interpret functions on the sphere as random variables. 
The expectation value of the Hamiltonian in a generic state
\begin{eqnarray}
|x\rangle = \sin\half\,
\theta\,\re^{{\rm i}\phi}|E_1\rangle +
  \cos\half\,\theta|E_2\rangle
   \end{eqnarray}
expressed in a normalized energy basis is given  by
\begin{eqnarray}
H(x) &=& \frac {\langle x|{\hat H}|x\rangle} {\langle x|x\rangle} 
\nonumber  \\ &=&  
E_1\sin^2\half\,\theta + E_2\cos^2\half\,\theta 
\nonumber  \\ &=&  \half(E_1 + E_2) + \half(E_2 - E_1) \cos\theta,
\end{eqnarray}
or $H(x) = \bar E + \omega \cos \theta$ in the more compact notation introduced above. 

Now we are in a position to work out the probability law for $H(x)$ under $\mathbb P$. 
Clearly $\mathbb{P} (H (x) < E_1) = 0$ and $\mathbb{P} (H (x) \leq E_2) = 1$. Then for 
$ E_1 \leq E \leq E_2$ we have
\begin{eqnarray}
\mathbb{P} (H \leq E) =  \frac {1} {4\pi}  \int_{\theta = \pi} ^{\theta_0}  \int_{\phi = 0} ^{2\pi} \sin \theta \, \rd \theta \, \rd \phi,
\end{eqnarray}
where 
\begin{eqnarray}
\theta_0 = \cos^{-1} \left( \frac {E - \bar E} {\omega} \right).
\end{eqnarray}
In other words, $\theta_0$ is the value of $\theta$ such that $\bar E + \omega \cos \theta = E$. 
The geometrical picture here is that the north and south poles of the sphere represent the two energy eigenstates, and the circles of constant latitude correspond to level values of the specific energy. The integral is then taken over the whole of that part of the surface of the sphere at and below the latitude corresponding to $E$. The integration is straightforward to carry out and we obtain
\begin{eqnarray}
\mathbb{P} (H \leq E) =  \frac {E - E_1} {E_2 - E_1},
\end{eqnarray}
as required. Thus we have shown, in the case of a two-dimensional system, that the Hamiltonian function can be interpreted as a random variable that is uniformly distributed over the interval $[E_1, E_2]$. As a consistency check one can verify that the variance of a uniformly distributed random variance over the interval indicated is indeed given by 
$\sigma^2 = \frac{1}{12}(E_2-E_1)^2$.

The fact that the Hamiltonian function is uniformly distributed for $r = 2$ can be used to work out the partition function (\ref{partition function}) in that case, and we obtain
\begin{eqnarray}
Z(\beta) = {\mathbb E}\left[\re^{-\beta H}\right]   =   
 \frac{1}{\beta (E_2 - E_1)} \left(\re^{-\beta E_1} -  \re^{-\beta E_2} \right).
\end{eqnarray}
As a consequence we find that the energy is given as a function of $\beta$ by 
\begin{eqnarray}
E(\beta) = \frac {1} {\beta} + \frac {E_1 \, \re^{-\beta E_1} - E_2 \, \re^{-\beta E_2} }
{ \re^{-\beta E_1} -  \re^{-\beta E_2} } , 
\label{uniform 2}
\end{eqnarray}
or equivalently
\begin{eqnarray}
E(\beta) = \frac {1} {\beta} + \half (E_1 + E_2) - \omega \coth (\beta \omega).
\label{uniform 3}
\end{eqnarray}
In this case we can use the Laurent expansion 
\begin{eqnarray}
\coth x = \frac {1}{x} + \tfrac {1}{3} x  + \tfrac{1}{45} x^3 + \cdots
\end{eqnarray}
to give us a Taylor expansion for the energy, and the result is
\begin{eqnarray}
E(\beta) =\half (E_1 + E_2) - \tfrac {1}{3} \beta \omega^2 + \tfrac{1}{45} \beta^3 \omega^4 + \cdots,
\label{uniform 4}
\end{eqnarray}
which differs from (\ref{Boltzmann 4}) even at first order. Likewise we can work out the heat capacity, and in this case we obtain
\begin{eqnarray}
C(\beta) = k_B \left(1 - \beta^2 \omega^2  \rm{csch}^2  (\beta \omega) \right),
\label{heat capacity uniform}
\end{eqnarray}
which is nonvanishing at zero temperature. 

\begin{acknowledgments}
{We are grateful to participants at the third Applied Geometric Mechanics network meeting on  Geometric Quantum Dynamics, Brunel University London (October 2014), the International Conference on Quantum Control, Cinvestav, Mexico City (October 2014), the International Workshop on Quantum Informatics, ITMO University, St Petersburg (November 2014), the Clarendon Laboratory, University of Oxford (May 2016), and the international workshop on Analytic and Algebraic Methods in Physics, Prague (June 2016)  for helpful comments. We are also grateful to O. Dahlsten, M. P. M\"uller and an anonymous referee for helpful comments. Part of this work was completed at the Aspen Center for Physics, which is supported by National Science Foundation grant PHY-1066293.
}
\end{acknowledgments}


\onecolumngrid

\vspace{2.0cm} 

\end{document}